\documentclass[12pt,a4paper]{article}

\usepackage[a4paper,margin=2.0cm]{geometry}

\usepackage{amsmath}
\usepackage{amssymb}
\usepackage{graphicx}
\usepackage{xcolor}

\newcommand{\etaeff}{\eta_{\mathrm{eff}}}
\newcommand{\Deff}{D_{\mathrm{eff}}}

\begin{document}

\title{Quantum Brownian motion in a temperature gradient
}

\author{
  R.~J.~S.~Afonso$^{1}$, 
  P.~V.~Paraguass\'u$^{2}$, 
  T.~Werlang$^{3}$
  and D.~Valente$^{4,3}$
  \\[2ex]
  \normalsize $^{1}$Instituto de F\'{\i}sica de S\~ao Carlos, Universidade de S\~ao Paulo,
  \normalsize S\~ao Carlos, SP, Brazil
  \\[0.8ex]
  \normalsize $^{2}$Departamento de F\'{\i}sica, Pontif\'{\i}cia Universidade Cat\'olica
  \\
  \normalsize do Rio de Janeiro, Rio de Janeiro, RJ, Brazil
    \\[0.8ex]
  \normalsize $^{3}$Universidade Federal de Mato Grosso,
  \normalsize Cuiab\'a, MT, Brazil
  \\[0.8ex]
  \normalsize $^{4}$Centro Brasileiro de Pesquisas F\'{\i}sicas,
  \normalsize Rio de Janeiro, RJ, Brazil
  \\[1.5ex]
}

\date{\today}

\maketitle

\begin{abstract}
When a classical Brownian particle is subjected to a temperature gradient it can undergo thermophoresis, i.e., the particle gets transported towards the colder regions of the environment, which bears relevant consequences to nonequilibrium self-organization.
Here, we investigate how the quantum Brownian motion is affected by a temperature gradient.
We employ a recently proposed generalized system-plus-reservoir model, where one assumes a continuum field of thermal baths.
As our main result, we obtain the quantum influence functional, within the Feynmann-Vernon path integral approach.
To test convergence, we derive an effective Langevin equation for the center of mass of the quantum wavepacket, and show that it coincides with the classical Langevin equation at high temperatures.
Both the fully quantum (low temperatures) and the semiclassical (high temperature) limits show signatures of quantum thermophoresis in the quantum Brownian motion, a so far unexplored effect.
\end{abstract}

\vspace{2ex}

\section{Introduction}
\label{sec:intro}

The quantum Brownian motion is a paradigmatic model for open-system dynamics, especially when the system is strongly coupled to the environment, as is usually the case for superconducting circuits \cite{breuer2002theory,weiss2012quantum,hanggi2005fundamental}. 
In the so called Caldeira-Leggett model, a particle is coupled linearly to a bath of non-interacting harmonic oscillators and, by means of the influence-functional method by Feynman and Vernon \cite{feynman1963theory}, dissipation and noise from a Hamiltonian universe can be deduced \cite{caldeira1983path,caldeira2014introduction,grabert1988quantum,ingold2002path}.
This agrees with the operator-based quantum Lagevin equation \cite{ford1988quantum}, and with the exact master equation for a general environment and a non-linear coupling \cite{hu1992quantum, hu1993quantum}. 
However, in all of this the environment is treated by a single temperature. 
Although multiple local temperatures do appear in models where a harmonic chain is used to study heat conduction \cite{rieder1967properties,dhar2008heat}, they are attached to the boudaries of the system rather than resolved as a field in space.
Alternativelly, models have been considered where  multiple particles are coupled to multiple thermal-equilibrium reservoirs \cite{mario10,mario16}.

A single particle in a {\it continuous} temperature gradient is a different problem. Thermophoresis, also known as the Soret effect ---~drift induced by a spatially varying temperature, known since 1856 \cite{ludwig1856diffusion,soret1879etat} and measured usually in colloidal and biomolecular suspensions \cite{duhr2006why,piazza2008thermophoresis,wurger2010thermal}~--- has no equilibrium state to expand around, and even classically its description depends on how the tempereature landscape influeces the noise. 
A classical Langevin equation with position-dependent diffusion might lead to different stochastic interpretations, with different steady states \cite{vankampen1988diffusion,lau2007state,sokolov2010ito}, and the drift induced by an inhomogeneous (multiplicative) noise intensity can be dominating
\cite{zochil2010,zochil2020,zochil2025}.
Experimentally, thermophoresis has gained renewed interest due to its consequences for nonequilibrium self-organization, remarkably in the context of the origin of biological cells on earth \cite{braun2025}.

Quantum thermophoresis has been recently uncovered with a single particle strongly confined (having very few energy levels), and weakly coupled to a discrete number of thermal baths \cite{mauricio2026thermoforesis}.
However, the quantum master equation employed in Ref.\cite{mauricio2026thermoforesis} cannot be applied to the free-particle limit, where the energy gaps vanish and the coupling to the environment becomes predominant, as is the case for quasi-free quantum Brownian particles.
The first step towards studying quantum thermophoresis for a quantum Brownian particle is to devise a generalized system-plus-reservoir Hamiltonian that describes a temperature gradient.
Such a Hamiltonian model has been recently introduced in Ref.\cite{valente2026thermoforesis}, here referred to as a generalized Caldeira--Leggett model (GCL).
The environment consists not of one single bath, but of a continuum of independent local baths, indexed by position $X$ and each in equilibrium at its own temperature, $T(X)$, coupled to the particle through a generic spatial profile.
Classically, GCL yields a Langevin equation with position-dependent friction and position-dependent noise (since the local noise depends on the local temperature), and leads to a thermophoretic (thermophobic) behavior.

Here, we go a step further so as to quantize the GCL of Ref.\cite{valente2026thermoforesis}.
Our direct quantization of a system-plus-reservoir Hamiltonian which itself originates a multiplicative noise is alternative to other methods such as departing from a Langevin dynamics \cite{zochil2010}.
We integrate the local baths out exactly along the closed time path \cite{schwinger1961brownian,keldysh1965diagram,calzetta2008nonequilibrium} and follow the consequences down to the classical limit.
Four results structure this paper. 
First (section~\ref{sec:kernels}), the exact influence functional splits into a dissipation kernel, which does not depend on the inverse temperature $\beta(X)$, and a noise kernel, that carries the whole temperature profile. 
The thermal gradient is a property of the fluctuations alone. Second (section~\ref{sec:langevin}), the semiclassical truncation gives a non-Markovian quantum Langevin equation for an arbitrary spectral density, an arbitrary coupling profile and an arbitrary smooth $T(X)$. 
Third (section~\ref{sec:ohmic}), the Ohmic limit gives us a local but position-dependent friction.
Fourth (section~\ref{sec:classical-limit}), the high-temperature limit reproduces the classical thermophoretic Langevin equation of Ref.\cite{valente2026thermoforesis} term by term. 

\section{The model and its exact influence functional}
\label{sec:model}

\subsection{The model}
\label{sec:model-H}

The GCL Hamiltonian is given by \cite{valente2026thermoforesis}
\begin{equation}
    H=\frac{p^2}{2M}+V(x)+\int dX \sum_k\left[\frac{p_k(X)^2}{2m_k}
    +\frac{m_k\omega_k^2}{2}\left(q_k(X)-\frac{c_k}{m_k\omega_k^2}\,x\,g(x-X)\right)^{2}\right],
    \label{eq:H-start}
\end{equation}
where the particle of mass $M$, position $x$ and momentum $p$, moves in a potential $V(x)$, and $X$ labels the position of a local bath. 
To be more precise, we shall integrate as $\int dX/L$, where $L$ has units of length, and set $L=1$.
It is convenient to define the coupling function
\begin{equation}
    h_X(x):=x\,g(x-X),
    \label{eq:hX-def}
\end{equation}
in terms of which \eqref{eq:H-start} reads
\begin{equation}
    H=H_S+\int dX \sum_k\left[\frac{p_k(X)^2}{2m_k}+\frac{m_k\omega_k^2}{2}q_k(X)^2 -c_k h_X(x)\,q_k(X)+\frac{c_k^2}{2m_k\omega_k^2}h_X(x)^2\right],
    \label{eq:H-expanded}
\end{equation}
with $H_S=p^2/2M+V(x)$. 
The last term is the usual counterterm, here inherited from the completed square in \eqref{eq:H-start}. 
It will cancel out a frequency-renormalizing contribution (see  section~\ref{sec:diss-sector}). The derivative of the coupling function,
\begin{equation}
    F(x,X):=\partial_x h_X(x)=g(x-X)+x\,g'(x-X),
    \label{eq:F-def}
\end{equation}
is the object that survives into every result below. 
It is also what appears in the classical treatment of the same Hamiltonian \cite{valente2026thermoforesis}. 
The form of $g$ decides which local baths the particle at $x$ is in contact with. 
It is assumed smooth and decaying at infinity, but is otherwise arbitrary. 
Note that the coupling \eqref{eq:hX-def} is nonlinear in the system coordinate, so that the model is not of the bilinear Caldeira--Leggett type even at a fixed $X$ \cite{hu1993quantum, paraguassu2026}. 
Our derivation in fact benefits from the linear coupling in the bath coordinates.

We should notice that the environment of \eqref{eq:H-start} is not a single heat-conducting medium. The bath Hamiltonian contains no term coupling oscillators at different $X$, so an oscillator at $X$ and one at $X'\neq X$ never interact and heat never flows through the environment itself; this is what distinguishes the model from a conducting medium such as a harmonic chain driven at its two ends \cite{rieder1967properties,dhar2008heat}. The temperature landscape is therefore inert on its own; it is explored only by the particle, which as it drifts from $x=X_1$ to $x=X_2$ exchanges energy with baths held at $T(X_1)$ and then at $T(X_2)$. 
The non-equilibrium character of the problem is carried entirely by the system. This is a modelling assumption and not a derivation --- in a real fluid the environment would have spatial correlations, $\langle q_k(X)q_{k'}(X')\rangle \neq 0$ \cite{de2006hydrodynamic}--- but it is exactly what makes an exact influence functional with a position-resolved temperature possible \cite{cavina2026quantum}. Moreover, one can assume that the fluid correlations occur on a timescale much shorter than that of the quantum Brownian particle's dynamics.

We take the total initial state to be factorized,
\begin{equation}
    \rho_{\mathrm{tot}}(0)=\rho_S(0)\otimes \rho_B(0),
    \label{eq:factorized}
\end{equation}
with the environment in a product of local thermal states,
\begin{equation}
    \rho_B(0)=\bigotimes_X \rho_{B,X},\qquad\rho_{B,X}=\frac{e^{-\beta(X) H_B(X)}}{Z_X},\qquad \beta(X)=\frac{1}{k_B T(X)}.
    \label{eq:local-thermal-state}
\end{equation}
Because the local baths are mutually independent, \eqref{eq:local-thermal-state} commutes with the free bath Hamiltonian and is stationary under the bath dynamics: each bath stays at its own temperature, and no relaxation of the landscape competes with the particle's dynamics. The reduced density matrix at time $t$ is
\begin{equation}
    \rho_S(x_f,x_f';t)=\int dx_i\,dx_i'\,J_r(x_f,x_f';t|x_i,x_i';0)\,\rho_S(x_i,x_i';0),
    \label{eq:reduced-density}
\end{equation}
with the reduced propagator
\begin{equation}
    J_r(x_f,x_f';t|x_i,x_i';0)=\int_{x_i}^{x_f}\mathcal D x\int_{x_i'}^{x_f'}\mathcal D x'\,\exp\!\left\{\frac{i}{\hbar}\bigl(S_S[x]-S_S[x']\bigr)\right\}\mathcal F[x,x'],
    \label{eq:Jr-def}
\end{equation}
where
\begin{equation}
    S_S[x]=\int_0^t ds\left[\frac{M}{2}\dot x^2-V(x)\right],
    \label{eq:Ssys}
\end{equation}
and every effect of the environment is contained in the influence functional $\mathcal F[x,x']$.

\subsection{Integrating out the local baths}
\label{sec:IF}

Each pair $(k,X)$ --- one oscillator mode of one local bath, henceforth simply a mode --- sees the particle as an external source. Writing
\begin{equation}
    J_{kX}(s):=c_k h_X(x(s)),\qquad J'_{kX}(s):=c_k h_X(x'(s)),
    \label{eq:J-sources}
\end{equation}
the action of one mode is
\begin{equation}
    S_{kX}[q;J]=\int_0^t ds\left[\frac{m_k}{2}\dot q^2-\frac{m_k\omega_k^2}{2}q^2+J_{kX}(s)q(s)\right],
    \label{eq:SkX}
\end{equation}
while the counterterm of \eqref{eq:H-expanded} contributes with a pure phase for the system,
\begin{equation}
    S_{\mathrm{CT}}[x]=-\int_0^t ds\int dX\sum_k\frac{c_k^2}{2m_k\omega_k^2}\,h_X(x(s))^2 .
    \label{eq:SCT-onebranch}
\end{equation}
Since the modes are independent and the initial state is \eqref{eq:local-thermal-state}, the influence functional factorizes,
\begin{equation}
    \mathcal F[x,x']=\exp\!\left\{\frac{i}{\hbar}\bigl(S_{\mathrm{CT}}[x]-S_{\mathrm{CT}}[x']\bigr)\right\}    \prod_X\prod_k\mathcal F_{kX}[J_{kX},J'_{kX}],
    \label{eq:F-factorized}
\end{equation}
with
\begin{align}
    \mathcal F_{kX}[J,J']&=\operatorname{Tr}\!\left[U_{J}(t)\,\rho_{B,X}^{(k)}\,U_{J'}^\dagger(t)\right]\nonumber\\
    &=\frac{1}{Z_{kX}}\int dq_i\,dq_i'\,dq_f\;\langle q_f|U_J(t)|q_i\rangle\,    \langle q_i|e^{-\beta(X)H_{B,k}}|q_i'\rangle\,\langle q_i'|U_{J'}^\dagger(t)|q_f\rangle .
    \label{eq:FkX-operator}
\end{align}
It is convenient to absorb the thermal weight into an imaginary-time branch, in the standard closed-time-path construction \cite{schwinger1961brownian,keldysh1965diagram,calzetta2008nonequilibrium,cavina2023convenient}. Denoting by $\gamma=\gamma_+\cup\gamma_-\cup\gamma_E$ the contour made of the forward and backward real-time branches and the Euclidean branch from $0$ to $-i\hbar\beta(X)$,
\begin{equation}
    \mathcal F_{kX}[J,J']=\frac{1}{Z_{kX}}\int\mathcal D q_C\,\exp\!\left\{\frac{i}{\hbar}S_{B,k}[q_C]+\frac{i}{\hbar}\int_{\gamma}dz\,J_{\gamma}(z)\,q_{\gamma}(z)\right\},
    \label{eq:FkX-contour}
\end{equation}
where $J_\gamma$ equals $J$ on $\gamma_+$, equals $J'$ on $\gamma_-$, and \emph{vanishes} on the Euclidean branch --- a direct consequence of the factorized initial state \eqref{eq:factorized}.

The integral is Gaussian and can be done exactly. Carrying out the Gaussian cumulant resummation (appendix~\ref{ap:cumulants}) gives, for one mode,
\begin{equation}
    \mathcal F_{kX}[J,J']=\exp\!\left[-\frac{1}{2\hbar^2}\int_\gamma dz\int_\gamma dz'\, J_\gamma(z)\,\langle\mathcal T_\gamma q_k(X,z)q_k(X,z')\rangle_{\beta(X)}\,J_\gamma(z')\right],
    \label{eq:FkX-exact}
\end{equation}
which is exact: the bath is Gaussian and centred, so all odd cumulants vanish and Wick's theorem \cite{wick1950evaluation,peskin2018introduction} resums the even ones into an exponential. Splitting the contour into its two real-time branches with
\begin{equation}
    G^{ab}_{kX}(s,s'):=\langle q_k^{\,a}(X,s)\,q_k^{\,b}(X,s')\rangle_{\beta(X)},\qquad a,b\in\{+,-\},
    \label{eq:Gab-def}
\end{equation}
and using $\int_\gamma dz\,J_\gamma q=\int_0^t ds\,[J q_+ - J' q_-]$, we obtain
\begin{align}
    \ln\mathcal F_{kX}[J,J']=-\frac{1}{2\hbar^2}\int_0^t\!\!ds\,ds'\Big[
    &J(s)G^{++}_{kX}(s,s')J(s')-J(s)G^{+-}_{kX}(s,s')J'(s')\nonumber\\
    -&J'(s)G^{-+}_{kX}(s,s')J(s')+J'(s)G^{--}_{kX}(s,s')J'(s')\Big].
    \label{eq:FV-version_1}
\end{align}
With $C^{>}(s,s')=\langle q(s)q(s')\rangle_\beta$ and $C^{<}(s,s')=\langle q(s')q(s)\rangle_\beta$, contour ordering gives $G^{++}=\theta(s-s')C^>+\theta(s'-s)C^<$, $G^{--}=\theta(s'-s)C^>+\theta(s-s')C^<$, $G^{-+}=C^>$ and $G^{+-}=C^<$.
Changing to the sum and difference sources
\begin{equation}
    \Sigma(s)=\frac{J(s)+J'(s)}{2},\qquad \Delta(s)=J(s)-J'(s),
    \label{eq:sigma-delta-def}
\end{equation}
the coefficient of $\Sigma\Sigma'$ cancels identically, the two mixed terms are equal after relabelling $s\leftrightarrow s'$, and one is left with (appendix~\ref{ap:sigma-delta})
\begin{align}
    \ln \mathcal F_{kX}[J,J']=-\Bigg[&\frac{1}{\hbar^2}\int_0^t\!\!ds\,ds'\,    \Delta(s)\,\theta(s-s')\big(C^>(s,s')-C^<(s,s')\big)\,\Sigma(s')\nonumber\\
    &+\frac{1}{4\hbar^2}\int_0^t\!\!ds\,ds'\,\Delta(s)\big(C^>(s,s')+C^<(s,s')\big)\Delta(s')\Bigg].
    \label{eq:FV-version_2}
\end{align}
The two brackets are the retarded and the symmetric bath propagators,
\begin{equation}
    \theta(s-s')\big(C^>(s,s')-C^<(s,s')\big)=\theta(s-s')\langle[q(s),q(s')]\rangle_\beta=-i\hbar\,G_R(s,s'),
    \label{eq:GR-identification}
\end{equation}
\begin{equation}
    C^>(s,s')+C^<(s,s')=\langle\{q(s),q(s')\}\rangle_\beta=G_S(s,s'),
    \label{eq:GS-identification}
\end{equation}
so that
\begin{equation}
    \ln\mathcal F_{kX}[J,J']=\frac{i}{\hbar}\int_0^t\!\!ds\,ds'\,\Delta(s)G_R(s,s')\Sigma(s')-\frac{1}{4\hbar^2}\int_0^t\!\!ds\,ds'\,\Delta(s)G_S(s,s')\Delta(s').
    \label{eq:FV-GR-GS}
\end{equation}
In terms of the paths, \eqref{eq:J-sources} and \eqref{eq:sigma-delta-def} give $\Delta(s)=c_k h_{\Delta,X}(s)$ and $\Sigma(s)=c_k h_{\Sigma,X}(s)$ with
\begin{equation}
    h_{\Delta,X}(s)=h_X(x(s))-h_X(x'(s)),\qquad h_{\Sigma,X}(s)=\tfrac12\big[h_X(x(s))+h_X(x'(s))\big],
    \label{eq:hDelta-hSigma}
\end{equation}
and the counterterm \eqref{eq:SCT-onebranch} becomes, using $h_X(x)^2-h_X(x')^2=2h_{\Delta,X}h_{\Sigma,X}$,
\begin{equation}
    S_{\mathrm{CT}}[x]-S_{\mathrm{CT}}[x']=-\int_0^t\!\!ds\int dX\sum_k    \frac{c_k^2}{m_k\omega_k^2}\,h_{\Delta,X}(s)\,h_{\Sigma,X}(s).
    \label{eq:SCT-sigma-delta}
\end{equation}

\subsection{The dissipation and the noise kernel}
\label{sec:kernels}

For a single oscillator of the local bath at $X$,
$q_k(X,s)=\sqrt{\hbar/2m_k\omega_k}\,(a_{kX}e^{-i\omega_ks}+a^\dagger_{kX}e^{i\omega_ks})$ with $\langle a^\dagger_{kX}a_{kX}\rangle_{\beta(X)}=n_k(X)=[e^{\beta(X)\hbar\omega_k}-1]^{-1}$,
so that
\begin{equation}
    C^>_{kX}(s,s')=\frac{\hbar}{2m_k\omega_k}\Big[(n_k(X)+1)e^{-i\omega_k(s-s')}+n_k(X)e^{i\omega_k(s-s')}\Big],
    \label{eq:Cgreater}
\end{equation}
and $C^<_{kX}(s,s')=C^>_{kX}(s',s)$. Hence, with $2n_k(X)+1=\coth[\beta(X)\hbar\omega_k/2]$,
\begin{equation}
    G_S(s,s')=\frac{\hbar}{m_k\omega_k}\coth\!\left(\frac{\beta(X)\hbar\omega_k}{2}\right)\cos\omega_k(s-s'), \qquad  G_R(s,s')=\theta(s-s')\frac{\sin\omega_k(s-s')}{m_k\omega_k}.
    \label{eq:GS-GR-explicit}
\end{equation}
Summing over the modes of the bath at $X$ defines the two kernels of the model,
\begin{equation}
    D_R(s,s')=\theta(s-s')\sum_k\frac{c_k^2}{m_k\omega_k}\sin\omega_k(s-s'),
    \label{eq:kernel-D_R}
\end{equation}
\begin{equation}
    D_S(X;s,s')=\sum_k\frac{\hbar c_k^2}{m_k\omega_k}    \coth\!\left(\frac{\beta(X)\hbar\omega_k}{2}\right)\cos\omega_k(s-s'),
    \label{eq:kernel-D_S}
\end{equation}
and assembling \eqref{eq:F-factorized}, \eqref{eq:FV-GR-GS} and
\eqref{eq:SCT-sigma-delta} gives the influence functional of the GCL,
\begin{align}
    \mathcal F[x,x']=\exp\Bigg\{&-\frac{i}{\hbar}\int_0^t\!\!ds\int dX\sum_k\frac{c_k^2}{m_k\omega_k^2}h_{\Delta,X}(s)h_{\Sigma,X}(s)\nonumber\\
    &+\frac{i}{\hbar}\int dX\int_0^t\!\!ds\int_0^t\!\!ds'\,h_{\Delta,X}(s)\,D_R(s,s')\,h_{\Sigma,X}(s')\nonumber\\
    &-\frac{1}{4\hbar^2}\int dX\int_0^t\!\!ds\int_0^t\!\!ds'\,h_{\Delta,X}(s)\,D_S(X;s,s')\,h_{\Delta,X}(s')\Bigg\}.
    \label{eq:FV-full}
\end{align}
Equation \eqref{eq:FV-full} is our main result.
First of all, it is exact. 
We did not use any expansion in the coupling, or any approximation on $T(X)$. 
The dissipation kernel, shown in Eq.\eqref{eq:kernel-D_R}, does not contain $\beta(X)$.
The noise kernel, on the other hand, as shown in Eq.\eqref{eq:kernel-D_S}, carries the temperature profile, through $\coth[\beta(X)\hbar\omega_k/2]$.
Quantum thermophoresis can be directly inferred from $D_S(X;s,s')$:
regions of higher temperatures have smaller $\beta(X)$, hence larger values for the diffusive term $D_S(X;s,s')$, thus larger tendency for the particle to escape region $X$.
When the particle finds lower temperatures, this escape tendency gets reduced, thus effectively trapping the particle.
In the classical GCL of Ref.\cite{valente2026thermoforesis}, the friction is also independent of the temperature, whereas thermal fluctuations depend on the local temperature. 
The following results are a consequence of this split.

\section{Semiclassical limit and the quantum Langevin equation}
\label{sec:semiclassical}

\subsection{Semiclassical paths and the dissipation sector}
\label{sec:diss-sector}

We introduce the mean coordinate, $x_c(s)$, associated with the semiclassical dyanmics and the difference coordinate, $y(s)$, being the coherence lenght \cite{weiss2012quantum} 
\begin{equation}
    x_c(s)=\frac{x(s)+x'(s)}{2},\qquad y(s)=x(s)-x'(s).
    \label{eq:xc-y-def}
\end{equation}
The kinetic part of \eqref{eq:Ssys} gives $\dot x^2-\dot x'^2=2\dot x_c\dot y$, and expanding the potential around $x_c$ gives $V(x)-V(x')\simeq yV'(x_c)$ to first order in $y$. Integrating the kinetic term by parts, with $y$ fixed at the endpoints,
\begin{equation}
    S_S[x]-S_S[x']\simeq-\int_0^t ds\,y(s)\big[M\ddot x_c(s)+V'(x_c(s))\big].
    \label{eq:Ssys-diff}
\end{equation}
The same expansion applied to \eqref{eq:hDelta-hSigma} gives
\begin{equation}
    h_{\Delta,X}(s)=y(s)F(x_c(s),X)+\mathcal{O}(y^3),\qquad
    h_{\Sigma,X}(s)=h_X(x_c(s))+\mathcal{O}(y^2),
    \label{eq:h-expansions}
\end{equation}
so that in every bilinear term appearing in \eqref{eq:FV-full} satisfies
\begin{equation}
    h_{\Delta,X}(s)h_{\Sigma,X}(s')\simeq y(s)\,F(x_c(s),X)\,h_X(x_c(s')).
    \label{eq:simple-product}
\end{equation}
The semiclassical approximation consists in keeping these leading orders; it is the statement that the off-diagonal extent $y$ of the density matrix is small compared with the scale on which $V$ and $g$ vary. That is, the coherence lenght $y$ is supressed by the semiclassical dynamics $x_c(s)$, i.e. $y(s)/x_c(s) \ll 1.$ The dynamics of the reduced system has no coherences. This is the standard truncation that turns an influence functional into a Langevin equation \cite{grabert1988quantum,calzetta2008nonequilibrium, paraguassu2026}, applied here with a coupling that is nonlinear in $x$.

Let us write $\ln\mathcal F=\ln\mathcal F_R+\ln\mathcal F_N$, with
\begin{align}
    \ln\mathcal F_R[x,x']=&-\frac{i}{\hbar}\int_0^t\!\!ds\int dX\sum_k\frac{c_k^2}{m_k\omega_k^2} h_{\Delta,X}(s)h_{\Sigma,X}(s)\nonumber\\
    &+\frac{i}{\hbar}\int dX\int_0^t\!\!ds\int_0^t\!\!ds'\,h_{\Delta,X}(s)D_R(s,s')h_{\Sigma,X}(s'),
    \label{eq:dissipation_FV}\\
    \ln\mathcal F_N[x,x']=&-\frac{1}{4\hbar^2}\int X\int_0^t\!\!ds\int_0^t\!\!ds'\, h_{\Delta,X}(s)D_S(X;s,s')h_{\Delta,X}(s').
    \label{eq:noise_FV}
\end{align}
Using \eqref{eq:simple-product} in \eqref{eq:dissipation_FV},
\begin{equation}
    \ln\mathcal F_R\simeq\frac{i}{\hbar}\int_0^t\!\!ds\int dX\,y(s)F(x_c(s),X)    \left[-\sum_k\frac{c_k^2}{m_k\omega_k^2}h_X(x_c(s))+\int_0^t\!\!ds'\,D_R(s,s')h_X(x_c(s'))\right].
    \label{eq:dissipation-DR-form}
\end{equation}
Introducing the damping function in Caldeira's convention
\cite{caldeira2014introduction},
\begin{equation}
    \Gamma(s-s')=\theta(s-s')\sum_k\frac{c^2_k}{m_k\omega^2_k}\cos\omega_k(s-s'),
    \label{eq:kernel-D_R_Gamma}
\end{equation}
and using $\delta(s-s')\cos\omega_k(s-s')=\delta(s-s')$, one has $\partial_s\Gamma(s-s')=\delta(s-s')\sum_k c_k^2/m_k\omega_k^2-D_R(s,s')$, i.e.
\begin{equation}
    D_R(s,s')=\delta(s-s')\sum_k\frac{c^2_k}{m_k\omega^2_k}-\partial_s\Gamma(s-s').
    \label{eq:DR-Gamma-relation}
\end{equation}
The instantaneous piece is exactly the counterterm: substituting \eqref{eq:DR-Gamma-relation} into \eqref{eq:dissipation-DR-form}, the two terms in the square bracket cancel and only the derivative of the damping function survives,
\begin{equation}
    \ln\mathcal F_R\simeq-\frac{i}{\hbar}\int_0^t\!\!ds\int dX\,y(s)F(x_c(s),X)
    \int_0^t\!\!ds'\,\partial_s\Gamma(s-s')\,h_X(x_c(s')).
    \label{eq:FR-after-cancellation}
\end{equation}
This is what the counterterm in \eqref{eq:H-start} is for: without it the particle would acquire a spurious frequency renormalization proportional to $\sum_k c_k^2/m_k\omega_k^2$. With $\partial_s\Gamma=-\partial_{s'}\Gamma$, integrating by parts and using $\Gamma(s-s')\propto\theta(s-s')$, which vanishes at $s'=t>s$, together with
\begin{equation}
    \frac{d}{ds'}h_X(x_c(s'))=\partial_xh_X(x_c(s'))\,\dot x_c(s')=F(x_c(s'),X)\,\dot x_c(s'),
    \label{eq:hX-derivative}
\end{equation}
gives $\ln\mathcal F_R=\ln\mathcal F_a+\ln\mathcal F_b$ with
\begin{equation}
    \ln\mathcal F_a=-\frac{i}{\hbar}\int_0^t\!\!ds\int_0^t\!\!ds'\int dX\,    y(s)F(x_c(s),X)\,\Gamma(s-s')\,F(x_c(s'),X)\,\dot x_c(s')
    \label{eq:Fa-def}
\end{equation}
and
\begin{equation}
    \ln\mathcal F_b=-\frac{i}{\hbar}\int_0^t\!\!ds\int dX\,y(s)F(x_c(s),X)\,\Gamma(s)\,h_X(x_c(0)).
    \label{eq:Fb-def}
\end{equation}
The boundary term \eqref{eq:Fb-def} depends on the initial position and describes a transient of the preparation. It is consistent with \eqref{eq:factorized} to set $h_X(x_c(0))=0$: the factorized initial state carries no system--bath correlation, which is the statement that the coupling is switched on at $s=0^+$. We drop $\ln\mathcal F_b$ from here on and explain this choice in appendix \ref{app:intial-position}.

\subsection{Noise sector and the quantum Langevin equation}
\label{sec:langevin}

With \eqref{eq:h-expansions}, the noise part \eqref{eq:noise_FV} becomes
\begin{equation}
    \ln\mathcal F_N\simeq-\frac{1}{4\hbar^2}\int_0^t\!\!ds\int_0^t\!\!ds'\,y(s)\,    \widetilde D_S(s,s')\,y(s'),
    \label{eq:FN-semiclassical}
\end{equation}
with the window-weighted noise kernel
\begin{equation}
    \widetilde D_S(s,s'):=\int dX\,F(x_c(s),X)\,D_S(X;s,s')\,F(x_c(s'),X).
    \label{eq:DS-tilde-def}
\end{equation}
This is a Gaussian weight in $y$, and it can be traded for a linear coupling to an auxiliary field by the functional Hubbard--Stratonovich identity \cite{stratonovich1957method,hubbard1959calculation}
\begin{equation}
    \exp\left(-\frac{1}{4\hbar^2}\int_0^t\!\!ds\int_0^t\!\!ds'\,y(s)\widetilde D_S(s,s')y(s')\right)=\int\mathcal D\xi\,P[\xi]\exp\left(\frac{i}{\hbar}\int ds\,\xi(s)y(s)\right),
    \label{eq:noise_FV_HS}
\end{equation}
the continuum version of the elementary Gaussian integral
\begin{equation}
    e^{-\frac{1}{2}Ky^2}=\frac{1}{\sqrt{2\pi K}}\int_{-\infty}^{\infty}d\xi\,     e^{-\frac{\xi^2}{2K}+i\xi y},\qquad K>0 .
    \label{eq:HS-elementary}
\end{equation}
The proof, including the rescaling of the auxiliary field that fixes its normalization in the continuum limit, is given in Appendix~\ref{app:HS-functional}. The field $\xi$ is a zero-mean Gaussian process with
\begin{equation}
    \langle\xi(s)\rangle=0,\qquad
    \langle\xi(s)\xi(s')\rangle=\frac12\widetilde D_S(s,s')
    =\frac12\int dX\,F(x_c(s),X)\,D_S(X;s,s')\,F(x_c(s'),X).
    \label{eq:xi-correlator}
\end{equation}
It carries the quantum statistics of the baths through the $\coth$ in \eqref{eq:kernel-D_S}, and it samples the temperature landscape through the window $F(x_c(s),X)$ centred on the particle's own position at time $s$. No assumption that $T(X)$ is slowly varying, or locally flat, has been made.

Following the convention of the classical treatment \cite{valente2026thermoforesis}, define the spectral function
\begin{equation}
    J(\omega):=\frac{\pi}{2}\sum_k\frac{c_k^2}{m_k\omega_k}\delta(\omega-\omega_k),
    \label{eq:J-def}
\end{equation}
so that $\sum_k f(\omega_k)c_k^2/m_k\omega_k=\frac{2}{\pi}\int_0^\infty d\omega\,J(\omega)f(\omega)$ for any test function $f$. Then \eqref{eq:kernel-D_R_Gamma} becomes a single spectral integral,
\begin{equation}
    \Gamma(s-s')=\theta(s-s')K(s-s'),\qquad K(\tau):=\frac{2}{\pi}\int_0^{\infty}d\omega\,\frac{J(\omega)}{\omega}\cos(\omega\tau),
    \label{eq:K-def}
\end{equation}
and $K$ carries no $X$: the spectral density is the same everywhere in space, only the occupation of the modes, i.e. $T(X)$, varies. This is the kernel-level statement of the split found in section~\ref{sec:kernels}.

Collect now every term of the exponent of \eqref{eq:Jr-def} that survives the semiclassical truncation: the system part \eqref{eq:Ssys-diff}, the dissipative part \eqref{eq:Fa-def}, and the noise part after \eqref{eq:noise_FV_HS}. All three are \emph{linear} in $y$, with coefficient
\begin{align}
    \Phi[x_c;s]=-\Big\{&M\ddot x_c(s)+V'(x_c(s))-\xi(s)\nonumber\\
    &+\int_0^s\!\!ds'\!\int dX\,F(x_c(s),X)\,\Gamma(s-s')\,F(x_c(s'),X)\,\dot x_c(s')\Big\}.
    \label{eq:Phi-def}
\end{align}
The remaining functional integral over $y$ is therefore a delta functional \cite{feynman2010quantum,kleinert2009path, paraguassu2026}: $y$ acts as a Lagrange multiplier enforcing $\Phi[x_c;s]=0$ for almost every $s$, that is,
\begin{equation}
    M\ddot x_c(s)+\int_0^s\!\!ds'\!\int dX\,F(x_c(s),X)\,\Gamma(s-s')\,F(x_c(s'),X)\,\dot x_c(s')
    =-V'(x_c(s))+\xi(s).
    \label{eq:general-langevin}
\end{equation}
This is the non-Markovian quantum Langevin equation of the GCL. It holds for an arbitrary spectral density $J(\omega)$, an arbitrary smooth coupling profile $g$, and an arbitrary smooth temperature landscape $T(X)$, which enters only through the statistics \eqref{eq:xi-correlator} of $\xi$. Both the friction kernel and the noise are non-local in time and non-local in space: the particle is damped by, and kicked by, every bath its window overlaps, weighted by $F(x_c,X)$.

\subsection{Ohmic regime}
\label{sec:ohmic}

For an Ohmic bath, $J(\omega)=\eta\omega$ with $\eta$ independent of position and temperature \cite{caldeira1983path,weiss2012quantum}, Eq.\eqref{eq:K-def} gives
\begin{equation}
    K(\tau)=\frac{2\eta}{\pi}\int_0^{\infty}d\omega\,\cos(\omega\tau)=2\eta\,\delta(\tau).
    \label{eq:K-ohmic}
\end{equation}
Substituting this directly into \eqref{eq:general-langevin} is not legitimate. The memory integral would contain
\begin{equation}
    \int_0^s\!\!ds'\,2\eta\,\delta(s-s')\,\theta(s-s')\,F(x_c(s'),X)\,\dot x_c(s'),
    \label{eq:theta-delta}
\end{equation}
in which the retarded step collapses onto the very point where the delta has its support. The product $\theta(0)\delta(0)$ is not defined: regularizing the step as $\theta_\epsilon$ and letting $\epsilon\to0$ gives $\theta(0)=0$, $\tfrac12$ or $1$ depending on the regularization, so the friction coefficient would be ambiguous by a factor of two. The ambiguity is an artefact of the order of operations, not a property of the model. In the semiclassical limit this ambiguity can be traced back to the stochastic Ito-Stratonovich preescription \cite{zochil2010, paraguassu2026}.
The retarded kernel \eqref{eq:kernel-D_R} is finite and unambiguous at any finite bandwidth; the Ohmic limit must therefore be taken in \eqref{eq:dissipation-DR-form}, before the $s'$ integration, and not afterwards. This is also the route taken in the classical treatment \cite{valente2026thermoforesis}.

Doing so, the square bracket of \eqref{eq:dissipation-DR-form} is, exactly,
\begin{equation}
    -\sum_k\frac{c_k^2}{m_k\omega_k^2}h_X(x_c(s))+\int_0^t\!\!ds'\,D_R(s,s')h_X(x_c(s'))
    =-\frac{d}{ds}\int_0^s\!\!ds'\,K(s-s')\,h_X(x_c(s')),
    \label{eq:DR-to-Kderivative}
\end{equation}
which displays the counterterm cancellation of section~\ref{sec:diss-sector} once more. For the Ohmic kernel \eqref{eq:K-ohmic} one now needs only the well-defined distributional identity
\begin{equation}
    \frac{d}{ds}\int_0^s\!\!ds'\,\delta(s-s')f(s')=\tfrac12\dot f(s)+f(0)\delta(s),
    \label{eq:half-derivative-identity}
\end{equation}
in which the factor $\tfrac12$ is fixed, not chosen: it is the value obtained from any symmetric regularization of the delta at finite bandwidth, and the same identity resolves the corresponding boundary in the classical derivation \cite{valente2026thermoforesis}. Hence
\begin{align}
    -\frac{d}{ds}\int_0^s\!\!ds'\,2\eta\,\delta(s-s')h_X(x_c(s')) &
    =-\eta\frac{d}{ds}h_X(x_c(s))-2\eta\,h_X(x_c(0))\delta(s)\nonumber\\
    &=-\eta\,F(x_c(s),X)\,\dot x_c(s),
    \label{eq:ohmic-boundary-identity}
\end{align}
using \eqref{eq:hX-derivative} and, as in \eqref{eq:Fb-def}, $h_X(x_c(0))=0$. The memory integral of \eqref{eq:general-langevin} therefore becomes local,
\begin{equation}
    \int_0^s\!\!ds'\!\int dX\,F(x_c(s),X)\Gamma(s-s')F(x_c(s'),X)\dot x_c(s')
    \;\xrightarrow{\ \text{Ohmic}\ }\;\etaeff[x_c(s)]\,\dot x_c(s),
    \label{eq:memory-to-local}
\end{equation}
with a friction coefficient that keeps the memory of the coupling window:
\begin{equation}
    \etaeff[x]:=\eta\int dX\,F(x,X)^2 .
    \label{eq:eta-eff-def}
\end{equation}
Note that $\dot x_c$ leaves the $X$ integral because \eqref{eq:ohmic-boundary-identity} has already collapsed the time integral onto $s'=s$; the remaining $X$ integral is smooth and gives the positive weight $F(x,X)^2$. Equation \eqref{eq:eta-eff-def} agrees term by term with the friction obtained from the classical equations of motion for the same Hamiltonian  \cite{valente2026thermoforesis}, from an entirely different technique.

Collecting, the semiclassical dynamics of the particle in the Ohmic GCL obeys the Markovian quantum Langevin equation
\begin{equation}
    \boxed{\;M\ddot x_c(t)+\etaeff[x_c(t)]\,\dot x_c(t)=-V'(x_c(t))+\xi(t)\;}
    \label{eq:markovian-langevin}
\end{equation}
with $\xi$ a zero-mean Gaussian noise whose correlator, for a completely general
smooth $T(X)$, is
\begin{equation}
    \langle\xi(t)\xi(t')\rangle=\frac12\int dX\,F(x_c(t),X)\,D_S(X;t,t')\,F(x_c(t'),X),
    \label{eq:noise-correlator-final}
\end{equation}
with $D_S$ given by \eqref{eq:kernel-D_S}. The friction has become local in time but remains a functional of position; the noise remains coloured, because the $\coth$ in $D_S$ decays over the thermal time $\hbar\beta(X)$ rather than collapsing to a delta.
Equations \eqref{eq:markovian-langevin}--\eqref{eq:noise-correlator-final} are the key results of this paper, as they guarantee the appropriate classical limit of our quantum treatment.

\subsection{The high-temperature limit}
\label{sec:classical-limit}

At high temperature, $k_BT(X)\gg\hbar\omega_k$ for all relevant modes,
$\coth[\beta(X)\hbar\omega_k/2]\to2/[\beta(X)\hbar\omega_k]$ and the noise
kernel \eqref{eq:kernel-D_S} collapses onto the damping kernel,
\begin{equation}
D_S(X;s,s')\to2k_BT(X)\sum_k\frac{c_k^2}{m_k\omega_k^2}\cos\omega_k(s-s')
    =2k_BT(X)K(s-s')=4\eta k_BT(X)\delta(s-s').
    \label{eq:DS-classical-limit}
\end{equation}
The correlator \eqref{eq:noise-correlator-final} then becomes white, with a position-dependent strength,
\begin{equation}
    \langle\xi(t)\xi(t')\rangle\to2\eta k_B\!\int\! dX\,F(x_c(t),X)^2T(X)\,\delta(t-t')
    \equiv2\Deff(x_c(t))\,\delta(t-t'),
    \label{eq:noise-correlator-classical-limit}
\end{equation}
which defines
\begin{equation}
    \Deff(x)=\eta k_B\int dX\,F(x,X)^2\,T(X).
    \label{eq:Deff-def}
\end{equation}
Equations \eqref{eq:markovian-langevin}, \eqref{eq:eta-eff-def} and \eqref{eq:noise-correlator-classical-limit} are precisely the classical thermophoretic Langevin equation of \cite{valente2026thermoforesis}, term by term: 
their Eq.~(48) is \eqref{eq:markovian-langevin}, their Eq.~(50) is \eqref{eq:eta-eff-def}, and their Eq.~(55) is \eqref{eq:noise-correlator-classical-limit}. 
The classical equation is thus not an input of the quantum theory but its high-temperature limit.

It is now worth analyzing the opposite, zero-temperature limit.
We notice that, for $\beta(X) \to \infty$, we find
$D_S(X;s,s') \to \hbar\sum_k\frac{c_k^2}{m_k\omega_k}\cos\omega_k(s-s') = \hbar K(s-s')$, but the Kernel computed with an {\it effective} spectral function 
$J_{\mathrm{eff}}(\omega) = J(\omega)\omega$.
In this sense, the deep quantum (low temperature) limit of the Ohmic regime behaves as a super-Ohmic classical (high temperature) limit, $J_{\mathrm{eff}}(\omega) = J(\omega)\omega = \eta \omega^2$.
In such an extreme limit, thermophoresis completely disappears, since no temperature dependence survives in the model.
From this, we hypothesize that gradients strong enough to cross the quantum-classical limit (from extremely low to classically high temperatures) should provide the most pronounced signatures of quantum thermophoresis.

To conclude the high-temperature limit, we add the classical Fokker-Planck equation that can be derived from Eqs.(\ref{eq:markovian-langevin})-(\ref{eq:Deff-def}) in the overdamped regime, $M\ddot x_c \to 0$, within the It\^o prescription \cite{valente2026thermoforesis},
\begin{equation}
\frac{\partial P(x,t)}{\partial t}
=
\frac{\partial}{\partial x}
\left[
\frac{V'(x)}{\eta_{\mathrm{eff}}(x)}P(x,t)
+
\frac{\partial}{\partial x}
\left[
\frac{D_{\mathrm{eff}}(x)}{\eta^2_{\mathrm{eff}(x)}}P(x,t)
\right]
\right],
\end{equation}
where $P(x,t)$ is the classical probability densitity of finding the Brownian particle at position $x$ at time $t$.
We do that to emphasize that this Fokker-Planck equation depends not only on $D_{\mathrm{eff}}(x)$ itself, but most importantly on
$\partial_x D_{\mathrm{eff}}(x)$.
In other words, thermophoresis depends crucially on the thermal gradient, going beyond the effective local temperature by itself.

\subsection{The sensed temperature}
\label{sec:Tbar}

By comparing Eqs.\eqref{eq:eta-eff-def} and \eqref{eq:Deff-def}, we see that the two coefficients differ only by the nonlocal thermal field $T(X)$, under the same positive
weight $F(x,X)^2$.
This suggests defining an effective local temperature, here termed the \emph{sensed temperature},
\begin{equation}
    \bar T(x):=\frac{\Deff(x)}{k_B\,\etaeff[x]}
    =\frac{\int dX\,F(x,X)^2\,T(X)}{\int dX\,F(x,X)^2},
    \label{eq:Tbar-def}
\end{equation}
a weighted average of the temperature landscape over the particle's own
coupling window.
The overlap between the functions $T(x)=T(X)|_{X=x}$ and $\bar T(x)$ can be regarded as a measure of how much local in space the thermal field is, thus providing us with a sense of volume for the otherwise pointlike (classical or quantum) particle.

\section{Conclusions and outlook}
\label{sec:conclusions}

In summary, we have quantized the generalized Caldeira--Leggett model (GCL) from Ref.\cite{valente2026thermoforesis} by integrating its continuum of independently thermalized local baths out of the closed-time-path path integral. 
The resulting influence functional \eqref{eq:FV-full} is exact and splits into a dissipation kernel that is independent on the temperature landscape, and a noise kernel that carries all of it.
From the influence functional, we have inferred the presence of thermophoresis.
The semiclassical truncation gives us the non-Markovian quantum Langevin equation \eqref{eq:general-langevin} for an arbitrary spectral density, an arbitrary coupling profile and an arbitrary smooth $T(X)$.
In the Ohmic case, with the limit taken before the time integration so that no ambiguous product of distributions arises, the friction becomes local but position dependent, Eqs.\eqref{eq:markovian-langevin}--\eqref{eq:eta-eff-def}, while the noise stays coloured. 
At high temperature the classical thermophoretic Langevin equation is recovered term by term.

We now aim at contrasting our direct quantization method with the indirect procedure quantizing the classical Langevin equation itself, as developed by Camurati and colleagues \cite{paraguassu2026}.
Understanding how temperature gradients affect quantum Brownian motion could be particularly useful to quantum optomechanical experiments with levitating spheres \cite{guerreiro2024,guerreiro2025,guerreiro2025paraguassu,guerreiro2026}.
This type of experiment could allow one to test our hypothesis, namely, that the most pronounced quantum thermophoresis would take place when the lowest temperatures of the non-homogeneous environment reached the quantum limit ($k_B T \ll \hbar \omega$, given a  harmonic frequency $\omega$) whereas the highest temperatures reached the classical limit ($k_B T \gg \hbar \omega$).

\section*{Acknowledgments}
R. J. S. A was financed by the São Paulo Research Foundation (FAPESP), Brasil, process Number 2025/15689-1.
D.V. was supported by CNPq (Grants 402074/2023-8; 408990/2025-2).
T. W. was supported by CNPq (Grants 402074/2023-8). P.V.P acknowledges the Funda\c{c}\~ao de Amparo \`a Pesquisa do Estado do Rio de Janeiro (FAPERJ Process SEI-260003/000174/2024).
\section*{Data availability statement}
No new data were created or analysed in this study. 

\appendix

\section{Gaussian cumulant evaluation of the single-mode functional}
\label{ap:cumulants}

We derive \eqref{eq:FkX-exact}. Omitting the mode indices, the object to compute is $F[J,J']=\operatorname{Tr}[U_J(t)\rho_BU^\dagger_{J'}(t)]$ with
\begin{equation}
    U_J(t)=\mathcal T\exp\left[-\frac{i}{\hbar}\int_0^t ds\,(H_0-J(s)q)\right],     \qquad H_0=\frac{p_k^2}{2m_k}+\frac{1}{2}m_k\omega_k^2q_k^2 . 
    \label{eq:UJ-def}
\end{equation}
Write $U_J(t)=U_0(t)U_I[J](t)$ with $U_0(t)=e^{-iH_0t/\hbar}$. From $i\hbar\,d_tU_J=(H_0-J(t)q)U_J$ and $i\hbar\,\partial_tU_0=H_0U_0$,
\begin{equation}
    U_0\,i\hbar\frac{dU_I[J]}{dt}=-J(t)\,q\,U_0U_I[J] \quad\Longrightarrow\quad
    i\hbar\frac{dU_I[J]}{dt}=-J(t)\,U_0^{\dagger}qU_0\,U_I[J],
    \label{eq:UI-eom}
\end{equation}
so that, with $q(t)=U_0^\dagger qU_0$ the free Heisenberg operator,
\begin{equation}
    U_J(t)=e^{-\frac{i}{\hbar}H_0t}\,\mathcal T\exp\left[\frac{i}{\hbar}\int_0^t ds\,J(s)q(s)\right].
    \label{eq:UJ-interaction}
\end{equation}
Substituting into $F[J,J']$, the free propagators cancel under the trace and
\begin{equation}
    F[J,J']=\Big\langle\widetilde{\mathcal T}e^{-\frac{i}{\hbar}\int_0^t ds\,J'(s)q(s)}\,\mathcal T e^{\frac{i}{\hbar}\int_0^t ds\,J(s)q(s)}\Big\rangle_\beta=\Big\langle\mathcal T_\gamma\exp\Big[\frac{i}{\hbar}\int_\gamma dz\,J_\gamma(z)q(z)\Big]\Big\rangle_\beta,
    \label{eq:F-contour-ordered}
\end{equation}
where the anti-time-ordered and time-ordered exponentials have been unified into a single contour-ordered one, the backward branch supplying the minus sign through $\int_\gamma dz\,J_\gamma q=\int_0^tds\,[Jq_+-J'q_-]$. 

Expanding \eqref{eq:F-contour-ordered} in powers of the source, the first-order term vanishes because the thermal state is centred, $\langle q\rangle_\beta=0$, and all odd terms vanish by the same parity argument: with $\Pi=e^{i\pi a^\dagger a}$ one has $\Pi a\Pi^\dagger=-a$, so any odd number of $q$'s has zero thermal average. At fourth order, Wick's theorem \cite{wick1950evaluation,peskin2018introduction} gives three identical pairings, 
\begin{equation}
    F^{(4)}=\frac{3}{4!}\left(\frac{i}{\hbar}\right)^4A[J_\gamma]^2
    =\frac{1}{2!}\left(-\frac{1}{2\hbar^2}A[J_\gamma]\right)^2,
    \qquad
    A[J_\gamma]=\int_\gamma\!\!dz\,dz'\,J_\gamma(z)\langle\mathcal T_\gamma q(z)q(z')\rangle_\beta J_\gamma(z'),
    \label{eq:F4}
\end{equation}
which is the third term of the series of $\exp[-A[J_\gamma]/2\hbar^2]$. The same argument at every even order --- Wick's theorem applies because the state is Gaussian --- resums the series into \eqref{eq:FkX-exact}.

\section{The $\Sigma/\Delta$ change of variables}
\label{ap:sigma-delta}

We derive \eqref{eq:FV-version_2} from \eqref{eq:FV-version_1}. Write the bracket of \eqref{eq:FV-version_1} as $\mathcal B=JG^{++}J'-JG^{+-}J'-J'G^{-+}J+J'G^{--}J'$, understood with the first factor at $s$ and the primed one at $s'$, and substitute $J=\Sigma+\tfrac{\Delta}{2}$, $J'=\Sigma-\tfrac{\Delta}{2}$. Collecting the four
bilinears,
\begin{align}
    \mathcal B=\;&\Sigma\big[G^{++}-G^{+-}-G^{-+}+G^{--}\big]\Sigma'    +\tfrac12\Sigma\big[G^{++}+G^{+-}-G^{-+}-G^{--}\big]\Delta'\nonumber\\
    &+\tfrac12\Delta\big[G^{++}-G^{+-}+G^{-+}-G^{--}\big]\Sigma'    +\tfrac14\Delta\big[G^{++}+G^{+-}+G^{-+}+G^{--}\big]\Delta'.
    \label{eq:B-four-terms}
\end{align}
Now insert $G^{++}=\theta C^>+(1-\theta)C^<$, $G^{--}=(1-\theta)C^>+\theta C^<$, $G^{-+}=C^>$ and $G^{+-}=C^<$, with $\theta\equiv\theta(s-s')$ and $1-\theta(s-s')=\theta(s'-s)$. The four coefficients become
\begin{equation}
    \begin{aligned}
    \Sigma\Sigma':&\quad (C^>+C^<)-(C^<+C^>)=0,\\
    \Sigma\Delta':&\quad \tfrac12\cdot2(\theta-1)(C^>-C^<)=(\theta-1)(C^>-C^<),\\
    \Delta\Sigma':&\quad \tfrac12\cdot2\theta(C^>-C^<)=\theta(C^>-C^<),\\
    \Delta\Delta':&\quad \tfrac14\cdot2(C^>+C^<)=\tfrac12(C^>+C^<).
    \end{aligned}
    \label{eq:four-coefficients}
\end{equation}
The vanishing of the first is the statement that the influence functional cannot depend on the mean path alone: a common phase on both branches is unobservable. The second and third are the same term. Relabelling $s\leftrightarrow s'$ in the $\Sigma\Delta'$ integral and using $C^<(s,s')=C^>(s',s)$,
\begin{equation}
    \int\!\!ds\,ds'\,\Sigma(s)(\theta(s-s')-1)\big(C^>-C^<\big)\Delta(s')
    =\int\!\!ds\,ds'\,\Delta(s)\,\theta(s-s')\big(C^>-C^<\big)\Sigma(s'),
    \label{eq:relabel}
\end{equation}
since $\theta(s'-s)-1=-\theta(s-s')$ and $C^>(s',s)-C^<(s',s)=-(C^>-C^<)(s,s')$, the two sign changes cancelling. The two mixed terms therefore \emph{add}, and
\begin{equation}
    \mathcal B\to2\,\Delta(s)\theta(s-s')\big(C^>-C^<\big)\Sigma(s')     +\tfrac12\Delta(s)\big(C^>+C^<\big)\Delta(s').
    \label{eq:B-collected}
\end{equation}
Multiplying by the prefactor $-1/2\hbar^2$ of \eqref{eq:FV-version_1} gives \eqref{eq:FV-version_2}. It is this factor of two that leaves the dissipative term with $1/\hbar^2$ while the noise term keeps $1/4\hbar^2$.

\section{The initial position issue}
\label{app:intial-position}
The integration by parts of the term inside \eqref{eq:FR-after-cancellation} is given by

\begin{equation}
    \int_0^t\!\!ds'\,\partial_s\Gamma(s-s')\,h_X(x_c(s')) = \Big[\Gamma(s-s')\,h_X(x_c(s'))\Big]_{s'=0}^{s'=t} - \int_0^t\!\!ds'\,\Gamma(s-s')\frac{d}{ds'}h_X(x_c(s')).
    \label{eq:D-initial-boundary-point}
\end{equation}
The upper limit vasinhes: $\Gamma(s-t)\propto \theta(s-t)=0$ for $s<t$ by \eqref{eq:kernel-D_R_Gamma}. The lower limit leaves $-\Gamma(s)h_X(x_c(0))$ and the remaining integral gives \eqref{eq:Fa-def} through \eqref{eq:hX-derivative}. The boundary term \eqref{eq:Fb-def} is what the integration by parts leaves behind at the initial time. It is entirely fixed by $x_c(0)$. By \eqref{eq:hX-def}, $h_X(x_c(0))=x_c(0)g(x_c(0)-X)$, which appears in \eqref{eq:Fb-def} under the whole landscape, weighted by $F(x_c(s),X)$. For any profile $g$ that is not identically zero, demanding the integrand vanish for every $X$ implies
\begin{equation}
    x_c(0)g(x_c(0)-X)=0 \qquad \text{for all}\quad X\quad \Longleftrightarrow \quad x_c(0)=0.
    \label{eq:D-x0-zero}
\end{equation}
The factorized preparation \eqref{eq:factorized} is the statement that the system-bath coupling is switched on abruptly at $s=0$. Let $f_{\varepsilon}$ be a smooth function, $f_{\varepsilon}(s)=0$ for $s\leq 0$ and $f_{\varepsilon}(s)=1$ for $s>\varepsilon$ and then replace $h_X(x_c(s))$ by $f_\varepsilon(s)h_X(x_c(s))$ throughout. We can reevaluate \eqref{eq:D-initial-boundary-point}. The first term on the right-hand side vanishes, however, the second term gives 

\begin{align}
    \int_0^t\!\!ds'\,\Gamma(s-s')\frac{d}{ds'}\left[f_\varepsilon(s')h_X(x_c(s'))\right]=&\int_0^t\!\!ds'\,\Gamma(s-s')\dot f_\varepsilon(s')h_X(x_c(s'))\notag\\
    &+\int_0^t\!\!ds'\Gamma(s-s')f_\varepsilon(s')\frac{d}{ds'}h_X(x_c(s'))
\end{align}
and $\dot f_\varepsilon(s') \rightarrow \delta(s')$ as $\varepsilon\rightarrow0$. Inserted into the second term of \eqref{eq:D-initial-boundary-point} we recover the boundary term. It is a signature of the abrupt switch-on, and it can be addressed with a statement about the initial condition.

Since \eqref{eq:Fb-def} is linear in $y$, keeping it should change nothing in the argument that leads from \eqref{eq:Phi-def} to \eqref{eq:general-langevin}, the functional integral over $y$ is still a delta functional and the Langevin equation acquired a new term called \emph{slip} force \cite{hanggi2005fundamental}.
That is,
\begin{equation}
        M\ddot x_c(s)+\int_0^s\!\!ds'\!\int dX\,F(x_c(s),X)\,\Gamma(s-s')\,F(x_c(s'),X)\,\dot x_c(s')=-V'(x_c(s))+\xi(s)+f_{\text{slip}}(s),
    \label{eq:D-langevin-with-slip}
\end{equation}
where
\begin{equation}
    f_{\text{slip}}(s):=-\Gamma(s)\int dXF(x,X)h_X(x_0)
\end{equation}
is a deterministic transient force. In the Ohmic regime this becomes negligible, as $\Gamma(s) \to \delta(s)$. 
Let us write the integral part of this force as
\begin{equation}
    \chi(x,x_0) := \int dXF(x,X)h_X(x_0).
    \label{eq:chi-def}
\end{equation}
We can write \eqref{eq:chi-def} in terms of $g$,
\begin{equation}
    \chi(x,x_0) = \int dX\left[x_0\,g(x-X)g(x_0-X)+x_0\,x\frac{\partial\,}{\partial x}g(x-X)g(x_0-X)\right].
\end{equation}
Taking $x=x_0$,
\begin{align}
    \chi(x_0,x_0) = \int dX\left[x_0\,g(x_0-X)^2+\,x_0^2\,g'(x_0-X)g(x_0-X)\right],
    \label{eq:chi-00}
\end{align}
where we have used $\partial_x\,g(x-X)|_{x=x_0}=g'(x_0-X)$. We should also notice $\partial_x\, g(x-X)=-\partial_X\,g(x-X)$, displacing the particle to a direction should have the same effect of moving the landscape in the opposite direction. We should also observe
\begin{equation}
    -g(x_0-X)\,\frac{\partial}{\partial X} g(x_0-X) = -\frac{1}{2}\frac{\partial}{\partial X}\left[g(x_0-X)\right]^2
\end{equation}
which lead us to change the second term in the right-hand side of \eqref{eq:chi-00},
\begin{equation}
    \int dX\,\left[\frac{\partial}{\partial x}g(x-X)\right]_{x=x_0}g(x_0-X) = -\frac{1}{2}\Big[g(x_0-X)^2\Big]_{X=-\infty}^{X=+\infty} = 0.
\end{equation}
Because $g$ decays to zero for baths that are very distant from the particle initial position, we can safely take the limit of the integral above; this leads us to
\begin{equation}
    \chi(x_0,x_0) = \int dXg(x_0-X)^2.
\end{equation}
However, since the domain of integration is not sensible to the particle finite value $x_0$ we can evaluate exactly at $x_0=0$, leading to
\begin{equation}
    \chi(x_0,x_0) = x_0 \int dXg(-X)^2 = \frac{\etaeff[0]}{\eta}x_0
\end{equation}
The last equality is derived from $F(0,X)=g(-X)$ by \eqref{eq:F-def} and \eqref{eq:eta-eff-def} in the ohmic regime. Hence, the friction coefficient at the decoupling points serves as a natural measure of the influence of the particle initial position. 
Alternatively, the presence of this initial impulse could also be eliminated by a different state preparation, with thermal coordinates shifted 
$q_k(X)\to q_k(X)-\big(c_k/m_k\omega_k^2\big)h_X(x_c(0))$ \cite{hanggi2005fundamental}.
The baths would then start already polarized by the particle. 
The slip force would have not arised due to the baths initial shift. 
But this would prevent us from writing the product form of Eq.\eqref{eq:local-thermal-state}, though.

\section{The functional Hubbard--Stratonovich identity}
\label{app:HS-functional}

Discretize the time integrals in the left-hand side of \eqref{eq:noise_FV_HS}, $s\to s_i$ with $i=1,\dots,N$ and $s_{i+1}-s_i=\Delta s$, and write $y_i=y(s_i)$:
\begin{equation}
    \exp\left(-\frac{1}{4\hbar^2}\int_0^t\!\!ds\int_0^t\!\!ds'\,y\widetilde D_Sy\right)
    \simeq\exp\left(-\frac12\sum_{i,j=1}^N y_i\,A_{ij}\,y_j\right),
    \qquad
    A_{ij}:=\frac{\Delta s^2}{2\hbar^2}\widetilde D_S(s_i,s_j).
    \label{eq:HS-discrete}
\end{equation}
$A$ is real, symmetric and positive definite, since $\widetilde D_S$ is a covariance. The elementary identity \eqref{eq:HS-elementary} in $N$ dimensions gives
\begin{equation}
    \exp\left(-\frac12\sum_{i,j}y_iA_{ij}y_j\right)
    =\frac{1}{\mathcal N}\int\prod_{i=1}^{N}d\xi_i\,
    \exp\left(-\frac12\sum_{i,j}\xi_i(A^{-1})_{ij}\xi_j+i\sum_i y_i\xi_i\right),
    \label{eq:discrete_multigaussian}
\end{equation}
with $\mathcal N=\sqrt{(2\pi)^N\det A}$, so that $\langle\xi_i\xi_j\rangle=A_{ij}$.
The continuum limit requires one further step, without which the source term and the covariance do not both survive. Define the rescaled field
\begin{equation}
    \xi(s_i):=\frac{\hbar}{\Delta s}\,\xi_i .
    \label{eq:xi-rescaling}
\end{equation}
Then the source term becomes $i\sum_iy_i\xi_i=\frac{i}{\hbar}\sum_i\Delta s\,y(s_i)\xi(s_i)\to \frac{i}{\hbar}\int ds\,\xi(s)y(s)$, reproducing the exponent of \eqref{eq:noise_FV_HS}, while the covariance becomes
\begin{equation}
    \langle\xi(s_i)\xi(s_j)\rangle=\frac{\hbar^2}{\Delta s^2}A_{ij}    =\frac{\hbar^2}{\Delta s^2}\cdot\frac{\Delta s^2}{2\hbar^2}\widetilde D_S(s_i,s_j)
    =\frac12\widetilde D_S(s_i,s_j),
    \label{eq:xi-covariance-discrete}
\end{equation}
in which every factor of $\hbar$ and $\Delta s$ cancels. Taking $\Delta s\to0$
gives \eqref{eq:noise_FV_HS} with the normalized Gaussian measure
\begin{equation}
    P[\xi]=\frac{1}{\mathcal N}\exp\left(-\frac12\int_0^t\!\!ds\int_0^t\!\!ds'\,    \xi(s)\left(\frac{\widetilde D_S(s,s')}{2}\right)^{-1}\xi(s')\right),  \qquad \int\mathcal D\xi\,P[\xi]=1,
    \label{eq:P-xi}
\end{equation}
and hence the moments quoted in \eqref{eq:xi-correlator}.

\bibliographystyle{unsrt}
\bibliography{bibliography}

\end{document}